\documentclass[apj,twocolumn]{emulateapj}

\usepackage{textcomp}
\usepackage{color}
\usepackage{graphicx}
\usepackage{tabularx}
\usepackage{natbib}
\usepackage{mathbbol}
\usepackage{amssymb}
\usepackage{amsmath}

\shorttitle{VHE gamma-ray observations of PSR J1023+0038}
\shortauthors{Aliu et al.}

\begin{document}

\title{A SEARCH FOR VERY HIGH-ENERGY GAMMA RAYS FROM THE MISSING LINK \\ 
 BINARY PULSAR J1023+0038 WITH VERITAS}

\author{
E.~Aliu\altaffilmark{1},
S.~Archambault\altaffilmark{2},
A.~Archer\altaffilmark{3},
W.~Benbow\altaffilmark{4},
R.~Bird\altaffilmark{5},
J.~Biteau\altaffilmark{6},
M.~Buchovecky\altaffilmark{7},
J.~H.~Buckley\altaffilmark{3},
V.~Bugaev\altaffilmark{3},
K.~Byrum\altaffilmark{8},
J.~V~Cardenzana\altaffilmark{9},
M.~Cerruti\altaffilmark{4},
X.~Chen\altaffilmark{10,11},
L.~Ciupik\altaffilmark{12},
M.~P.~Connolly\altaffilmark{13},
W.~Cui\altaffilmark{14},
H.~J.~Dickinson\altaffilmark{9},
J.~D.~Eisch\altaffilmark{9},
A.~Falcone\altaffilmark{15},
Q.~Feng\altaffilmark{14},
J.~P.~Finley\altaffilmark{14},
H.~Fleischhack\altaffilmark{11},
A.~Flinders\altaffilmark{16},
P.~Fortin\altaffilmark{4},
L.~Fortson\altaffilmark{17},
A.~Furniss\altaffilmark{18},
G.~H.~Gillanders\altaffilmark{13},
S.~Griffin\altaffilmark{2},
J.~Grube\altaffilmark{12},
G.~Gyuk\altaffilmark{12},
M.~H{\"u}tten\altaffilmark{11},
N.~H{\aa}kansson\altaffilmark{10},
J.~Holder\altaffilmark{19},
T.~B.~Humensky\altaffilmark{20},
C.~A.~Johnson\altaffilmark{6},
P.~Kaaret\altaffilmark{21},
P.~Kar\altaffilmark{16},
N.~Kelley-Hoskins\altaffilmark{11},
M.~Kertzman\altaffilmark{22},
D.~Kieda\altaffilmark{16},
M.~Krause\altaffilmark{11},
M.~J.~Lang\altaffilmark{13},
A.~Loo\altaffilmark{20},
G.~Maier\altaffilmark{11},
S.~McArthur\altaffilmark{14},
A.~McCann\altaffilmark{2},
K.~Meagher\altaffilmark{23},
P.~Moriarty\altaffilmark{13},
R.~Mukherjee\altaffilmark{1},
T.~Nguyen\altaffilmark{23},
D.~Nieto\altaffilmark{20},
A.~O'Faol\'{a}in de Bhr\'{o}ithe\altaffilmark{11},
R.~A.~Ong\altaffilmark{7},
A.~N.~Otte\altaffilmark{23},
D.~Pandel\altaffilmark{24},
N.~Park\altaffilmark{25},
V.~Pelassa\altaffilmark{4},
A.~Petrashyk\altaffilmark{20},
M.~Pohl\altaffilmark{10,11},
A.~Popkow\altaffilmark{7},
E.~Pueschel\altaffilmark{5},
J.~Quinn\altaffilmark{5},
K.~Ragan\altaffilmark{2},
P.~T.~Reynolds\altaffilmark{26},
G.~T.~Richards\altaffilmark{23},
E.~Roache\altaffilmark{4},
C.~Rulten\altaffilmark{17},
M.~Santander\altaffilmark{1},
G.~H.~Sembroski\altaffilmark{14},
K.~Shahinyan\altaffilmark{17},
A.~W.~Smith\altaffilmark{27},
D.~Staszak\altaffilmark{25},
I.~Telezhinsky\altaffilmark{10,11},
J.~V.~Tucci\altaffilmark{14},
J.~Tyler\altaffilmark{2},
A.~Varlotta\altaffilmark{14},
S.~Vincent\altaffilmark{11},
S.~P.~Wakely\altaffilmark{25},
O.~M.~Weiner\altaffilmark{20},
A.~Weinstein\altaffilmark{9},
A.~Wilhelm\altaffilmark{10,11},
D.~A.~Williams\altaffilmark{6},
B.~Zitzer\altaffilmark{8}, \\
and \\
M.~Chernyakova\altaffilmark{28,29},
M.~Roberts\altaffilmark{30,31}
}

\affil{$^{1}$Department of Physics and Astronomy, Barnard College, Columbia University, NY
10027, USA; {\color{blue}ester.aliu.fuste@gmail.com}}
\affil{$^{2}$Physics Department, McGill University, Montreal, QC H3A 2T8, Canada}
\affil{$^{3}$Department of Physics, Washington University, St. Louis, MO 63130, USA}
\affil{$^{4}$Fred Lawrence Whipple Observatory, Harvard-Smithsonian Center for
Astrophysics, Amado, AZ 85645, USA}
\affil{$^{5}$School of Physics, University College Dublin, Belfield, Dublin 4, Ireland}
\affil{$^{6}$Santa Cruz Institute for Particle Physics and Department of Physics,
University of California, Santa Cruz, CA 95064, USA}
\affil{$^{7}$Department of Physics and Astronomy, University of California, Los Angeles,
CA 90095, USA}
\affil{$^{8}$Argonne National Laboratory, 9700 S. Cass Avenue, Argonne, IL 60439, USA}
\affil{$^{9}$Department of Physics and Astronomy, Iowa State University, Ames, IA 50011,
USA}
\affil{$^{10}$Institute of Physics and Astronomy, University of Potsdam, 14476
Potsdam-Golm, Germany}
\affil{$^{11}$DESY, Platanenallee 6, 15738 Zeuthen, Germany}
\affil{$^{12}$Astronomy Department, Adler Planetarium and Astronomy Museum, Chicago, IL
60605, USA}
\affil{$^{13}$School of Physics, National University of Ireland Galway, University Road,
Galway, Ireland}
\affil{$^{14}$Department of Physics and Astronomy, Purdue University, West Lafayette, IN
47907, USA}
\affil{$^{15}$Department of Astronomy and Astrophysics, 525 Davey Lab, Pennsylvania State
University, University Park, PA 16802, USA}
\affil{$^{16}$Department of Physics and Astronomy, University of Utah, Salt Lake City, UT
84112, USA}
\affil{$^{17}$School of Physics and Astronomy, University of Minnesota, Minneapolis, MN
55455, USA}
\affil{$^{18}$Department of Physics, California State University - East Bay, Hayward, CA
94542, USA}
\affil{$^{19}$Department of Physics and Astronomy and the Bartol Research Institute,
University of Delaware, Newark, DE 19716, USA}
\affil{$^{20}$Physics Department, Columbia University, New York, NY 10027, USA}
\affil{$^{21}$Department of Physics and Astronomy, University of Iowa, Van Allen Hall,
Iowa City, IA 52242, USA}
\affil{$^{22}$Department of Physics and Astronomy, DePauw University, Greencastle, IN
46135-0037, USA}
\affil{$^{23}$School of Physics and Center for Relativistic Astrophysics, Georgia
Institute of Technology, 837 State Street NW, Atlanta, GA 30332-0430; {\color{blue}
gtrichards@gatech.edu}}
\affil{$^{24}$Department of Physics, Grand Valley State University, Allendale, MI 49401,
USA}
\affil{$^{25}$Enrico Fermi Institute, University of Chicago, Chicago, IL 60637, USA}
\affil{$^{26}$Department of Physical Sciences, Cork Institute of Technology, Bishopstown,
Cork, Ireland}
\affil{$^{27}$University of Maryland, College Park / NASA GSFC, College Park, MD 20742,
USA}
\affil{$^{28}$School of Physical Sciences, Dublin City University, Dublin 9, Ireland;
{\color{blue}masha.chernyakova@dcu.ie}}
\affil{$^{29}$Dublin Institute for Advanced Studies, 31 Fitzwilliam Place, Dublin 2,
Ireland}
\affil{$^{30}$New York University Abu Dhabi, P.O. Box 129188 Saadiyat Island, Abu
Dhabi, UAE; {\color{blue}malloryr@gmail.com}}
\affil{$^{31}$Eureka Scientific 2452 Delmer Street Suite 100 Oakland,
CA 94602}


\begin{abstract}
The binary millisecond radio pulsar PSR J1023+0038 exhibits many characteristics similar
to the gamma-ray binary system PSR B1259--63/LS 2883, making it an
ideal candidate for the study of high-energy non-thermal emission. It has been 
the subject of multi-wavelength campaigns following the disappearance of the pulsed
radio emission 
in 2013 June, which revealed the appearance of an accretion disk
around the neutron star. We present the results of very high-energy gamma-ray
observations carried out by VERITAS before and after this change of state. Searches for
steady and pulsed emission of both data sets yield no significant gamma-ray signal above
100\,GeV, and upper limits are given for both a steady and pulsed gamma-ray flux.  These
upper limits are used to constrain the magnetic field strength in the shock region of the
PSR J1023+0038 system.  Assuming that very high-energy gamma rays are produced via an 
inverse-Compton mechanism in the shock region, we constrain the shock magnetic field to be greater 
than $\sim$2\,G before the disappearance of the radio pulsar and greater than $\sim$10\,G afterwards.
\end{abstract}

\keywords{pulsars: general --- pulsars: individual(PSR J1023+0038) --- gamma
rays: general --- binaries: general}

\section{Introduction}
%%%%% Paragraph #1
Radio millisecond pulsars (MSPs) are old neutron stars that have been spun up to
millisecond periods via accretion of material from a companion star in a low-mass X-ray
binary~\citep[LMXB;][]{alpar1982}. In the past few years, new MSP discoveries have
taken place at a greatly elevated rate due to searches for radio pulsars in
unassociated {\it Fermi}-LAT-detected gamma-ray sources~\citep{paulray2012}.  This
new population of MSPs has enriched the known diversity of binary MSP companions.
This is especially true for eclipsing systems, which were rarely seen
outside of globular clusters: the ``black widows'' with very low-mass
(M $\ll$ 0.1M{$_{\odot}$) companions and ``redback'' systems with more massive (M$_c
\gtrsim$ 0.1M{$_{\odot}$), non-degenerate companions~\citep{roberts2011}.
Some of these redbacks have been observed to
transition between LMXB and MSP states, providing the first direct observational evidence
to support the theory of the MSP formation mechanism.
There are now three systems where transitions have been
observed: PSR J1023+0038~\citep{archibald2009} and XSS
12270--4859~\citep{bassa2014,roy2015} in the Galactic plane, and PSR
J1824--2452I~\citep{papitto2013}, located in the globular cluster M28.  Additionally, it
has recently been suggested that the galactic binary 1RXS J154439.4--112820 may
also be a transitional system~\citep{2015arXiv150805844B}.

%%%%% Paragraph #2
Very high-energy (VHE) gamma-ray emission ($E > 100$\,GeV) from binaries
containing MSPs has been
predicted to occur through diverse mechanisms. \cite{harding05} propose
that leptons
accelerated above the polar cap can produce inverse-Compton or curvature radiation
emission that could potentially be identified as gamma-ray pulsations at energies up to
 and above 100\,GeV, similar to what has been observed from the young Crab
pulsar~\citep{aliu2008,aliu2011}. Additionally, leptons could be accelerated at the shock
that appears as a result of the interaction between the pulsar wind and material ablated
off of the companion. These leptons could then radiate VHE gamma rays via 
inverse-Compton scattering, which could be modulated with the binary orbital period. This
emission scenario is thought to occur
in the VHE-detected binary system PSR B1259--63/LS 2883, a radio pulsar in a $\sim$3.4\,yr
orbit
around a
massive, luminous Be star~\citep{aharonian2005}. 

%%%%%%%%%%%%%%%%%%%%%%%%%%%%
\begin{table*}
\centering
\begin{tabular}{c c c c c c c c}
  \hline
  Binary State       &     On      &        Off      &     $\alpha$  &   Excess   &  LiMa 
           &    95\% CL flux UL  &  95\% CL flux UL (flux units)                     \\
                             &   Events  &    Events   &                     &   Events   
& Significance  &    (cm$^{-2}$s$^{-1}$)  & (erg cm$^{-2}$s$^{-1}$) \\ \hline
 Radio MSP          &     287     &     1815     &        0.17      &    -15.5      &    
-0.8           &     8.1 $\times 10^{-13}$   &  $5.8 \times 10^{-13}$                                   \\
 Accretion/LMXB   &     72      &        422     &        0.17      &       1.7      &    
 0.2           &    9.6 $\times 10^{-13}$   &  $6.9 \times 10^{-13}$                                    \\
\hline
\end{tabular}
\caption{\normalfont{VERITAS analysis results for the location of PSR J1023+0038 for each
of the two different 
binary states. The parameter $\alpha$ indicates the ratio of the on- to
off-source region exposure, and the LiMa significance is calculated using equation 17
in~\cite{lima1983}.}}
\label{tab:steadyfluxtable}
\end{table*}
%%%%%%%%%%%%%%%%%%%%%%%%%%%%

The theory of
VHE gamma-ray emission from PSR B1259--63~\citep{tavani1994} was first explored in the context of
the
original Black Widow Pulsar system~\citep{arons1993}, which is a binary comprising
the 1.6\,ms pulsar PSR B1957+20 in a 9.2\,hr orbit around a
$\sim$0.02 $M_{\odot}$ companion. However, no VHE
emission has been detected from the Black
Widow~\citep{otte2007}.  Searches for VHE emission from several globular clusters have
been undertaken,
since they are known to contain many of these 
eclipsing binary systems.  Recently, H.E.S.S. has detected VHE emission from the direction
of 
the cluster Terzan 5~\citep{abramowski2011}, which is especially rich in eclipsing
binary systems among globular clusters~\citep{2008IAUS..246..291R}.  This emission is
thought to originate in a 
bow shock region where interaction between leptons from MSP winds and the galactic medium 
occur~\citep{2014MNRAS.445.2842B}.  However, searches for VHE emission from the globular clusters  
47 Tuc~\citep{aharonian2009}, M5, M15 (McCutcheon et al. 2009),
%~\citep{2009arXiv0907.4974M}
 and M13~\citep{anderhub2009,
2012PhDT.......343M} have revealed no such emission.
The aforementioned eclipsing binary systems in globular clusters can be seen as
smaller-scale versions
of PSR B1259--63, because their more massive, nearly Roche-lobe-filling companions provide
much larger targets and more copious seed photons for inverse-Compton scattering than
companions of black widows. With the discovery of nearby redbacks in the Galactic
field, it is thought that a single, energetic Galactic-field redback could be  brighter at
VHE energies than the combined emission from many eclipsing systems in a distant
cluster~\citep{roberts2011}.

PSR J1023+0038 is a redback system containing a 1.69\,ms MSP in a
4.8\,hr orbit around a G star with a mass of $\sim$0.2M$_{\odot}$~\citep{archibald2009}.
PSR J1023+0038 was
selected as a promising candidate for VHE observations with VERITAS based on three
parameters thought to be responsible for the VHE emission from PSR B1259--63: the high
spin-down
luminosity of the
pulsar, the presence of an intense target photon field for inverse-Compton scattering
provided by the companion, and the relatively small distance from Earth. 
Although the optical
luminosity of the companion in PSR J1023+0038 is a factor of $\sim$10$^{4}$ less than that
for the
companion of PSR B1259--63, this discrepancy is possibly compensated by the much smaller
distance separating the pulsar and its companion in PSR J1023+0038, potentially making the
energy density of seed photons at the shock comparable for the two systems.  However,
the PSR B1259--63 system has a circumstellar disk that the pulsar passes through at
periastron~\citep{1998MNRAS.298..997W}, though PSR J1023+0038 shows no evidence
of such a disk. 

While the actual VHE emission will depend on the details of the flow
and the  magnetic field at the shock, the inverse-Compton emission should roughly
scale as $F_{{\rm IC}} \propto f (\dot E/d^2)u_{{\rm ph}}$ where $d$ is the distance 
to the binary, $u_{{\rm ph}} \sim
(R_c/R_s)^2 \sigma T_c^4/c$ is the photon energy density at the shock,
$R_c$ is the radius of the companion, $R_s$ is the radius of the shock
measured from the companion, and $f$ is a geometrical factor
representing the fraction of the pulsar wind involved in the
shock. If the shock region of PSR J1023+0038 (and by extension other redbacks
and black widows) is very near the surface of the companion, as
proposed by~\cite{bogdanov2011}, then $R_c/R_s \sim 1$, and $f$ is related to the
angle subtended by the companion in the pulsar sky. In the extreme
case of a shock only near the surface of the companion, $f$ is approximately $0.01$ % changed from 0.013 for sig figs
if the pulsar wind is isotropic, and $f$ is approximately $0.07$ if the wind is confined to
the equatorial plane.   Based on this simple estimation, the expected
 TeV flux from PSR J1023+0038 would be on the order of $\sim$\,0.1$f$
that of PSR B1259--63 near periastron, where it has an
observed flux $F(E>1{\rm TeV})\sim 10^{-11}{\rm cm}^{-2}\,{\rm s}^{-1}$~\citep{2013A&A...551A..94H}. 
We note that PSR J1023+0038 was selected for observations
before the publication of the revised estimates for the distance and spin-down power %from the VLBI measuements 
given in~\cite{deller2015}, in which case the estimated TeV flux would have been closer to $1f$ that of
PSR B1259--63.

Orbitally modulated X-ray emission has been
observed from PSR J1023+0038, suggesting that the system contains shocked
material~\citep{archibald2010}, and the observed radio eclipses suggest that the shock region
may be quite large.~\cite{tam10} found strong evidence of gamma-ray emission from the
direction of PSR J1023+0038 in the high-energy (HE; $100$\,MeV $\lesssim$ $E \leq
100$\,GeV) gamma-ray band
using {\it Fermi}-LAT data. Given the
observed
steep spectrum of this emission ($\Gamma\sim3$), the authors suggest that the gamma
rays likely originate from the pulsar magnetosphere rather than the intrabinary
shock.  Indeed,~\cite{archibald2013} have reported a hint of pulsed HE gamma-ray
emission from the pulsar magnetosphere with a statistical significance of $3.7\sigma$.

A sudden change of state in PSR J1023+0038 was reported to have occurred in 2013 June
after 
the pulsed radio emission from the MSP was no longer detected~\citep{stappers2013}, and
optical evidence for an accretion disk in the
system was found for
the first time since 2001 December~\citep{halpern2013, 2003AJ....126.1499S}. The
X-ray emission increased only moderately~\citep{kong2013, patruno2013}, implying that
accretion may still be inhibited due to the influence of the pulsar magnetosphere,
although low-level X-ray pulsations, thought to be powered by accretion, have been
detected~\citep{archibald2014}.  All of this new behavior coincided with a five-fold
increase in
the HE gamma-ray flux from PSR J1023+0038~\citep{2014ApJ...790...39S}. 

The similarities between PSR J1023+0038 and PSR B1259--63/LS 2883 motivated the first
VERITAS
observations of the PSR J1023+0038 in 2010.  
%in order investigate possible intrabinary shock
%emission from this type of system. 
Follow-up observations took place after the system was
reported to have transitioned to an accretion/LMXB state in 2013, prompted by the
substantial increase in flux in the HE gamma-ray band observed with the {\it Fermi}-LAT.
Here we report the results of these observations of PSR J1023+0038, the first ever made in the
VHE band, covering the two different states of this exceptional transitional object. After
describing the observations (\S~2), the analysis and results are presented (\S~3),
including searches for steady emission from the binary and pulsed emission from the
pulsar magnetosphere. In the final section we provide a simple spectral model to
interpret and discuss the results (\S~4).

\section{VHE gamma-ray observations}

The Very Energetic Radiation Imaging Telescope Array System (VERITAS) is a ground-based
array of four imaging atmospheric Cherenkov telescopes operating at the Fred Lawrence
Whipple Observatory (FLWO) in southern Arizona, USA. It was designed to explore the
universe in VHE gamma rays in the energy range from $\sim$85 GeV to above 30 TeV. 
Brief
Cherenkov light flashes from gamma-ray- and cosmic-ray-initiated air showers are focused
by 12\,m
reflectors onto cameras comprising 499 photomultiplier tubes located in the focal plane
of
each telescope, giving a $3^{\circ}.5$ diameter field of view. The angular resolution of
the array (68\% containment) reaches 0$^\circ$.08 per gamma-ray photon candidate, with a
sensitivity
to detect a point source at the 5$\sigma$ level with 1\% of the Crab Nebula flux above 300
GeV and at a 20$^{\circ}$ zenith angle in approximately 25\,hr. For a review of the
detector, see~\cite{holder2006, holder2008}.

The first observations of PSR J1023+0038 were made by VERITAS between 2010
December 8 and 2011 February 25 when the system was in the radio MSP state, resulting in 20\,hr
of live time available for analysis after data quality selection. Further observations
took place in 2013 December for a total 10\,hr of live time coinciding with the newly
reported accretion/LMXB state of the system. The two sets of data were recorded in two
different configurations of the VERITAS array: 2009 August to 2012 July, with the
original cameras and electronics; and 2012 August to present, following an upgrade to
the telescope cameras and the trigger system~\citep[for details, see][]{kieda2013}. Data
were taken on clear and moonless nights in {\it wobble} observation mode in which the
telescope pointing is offset by $0^{\circ}.5$ from the position of PSR J1023+0038, 
alternating between the four
cardinal directions to allow simultaneous accumulation of data and
background~\citep{fomin1994}. The data span the zenith angle range of $31^\circ$ to
$39^\circ$.

\section{VERITAS analysis \& results}

\subsection{Analysis} \label{sec:analysis}

The data are analyzed with a standard VERITAS software pipeline for reconstructing the
primary parameters of the gamma rays~\citep[see, for example,][]{acciari2008} and
cross-checked using an independent calibration and analysis software
package~\citep{cogan2008}.  The images are first flat-fielded by recording the response of
the cameras to a pulsed UV LED system~\citep{hanna2010}. The images are cleaned to remove
contamination by night sky background~\citep{daniel2008} and parameterized by their
principal moments~\citep{hillas1985}.  Arrival directions and impact distances are
estimated by analyzing the orientation of shower images in different
telescopes~\citep{hofmann1999}.  Background estimation is achieved by comparison of the
parameters of the recorded events with those computed in Monte Carlo gamma-ray
simulations~\citep{krawczynski2006}.

\subsection{Search for a steady signal}

A search for a VHE gamma-ray excess signal from the direction of PSR J1023+0038 is carried
out independently for 
the two states of the system observed with VERITAS. None of these searches yield a significant excess over the 
estimated background from the location of PSR J1023+0038.
Upper limits (ULs) on the integral flux above 300\,GeV from PSR J1023+0038 for each state
are set. The approach 
of~\cite{rolke2005} is used to determine these ULs at the 95\% confidence level (CL) 
assuming a power-law source 
spectrum with a photon index of $\Gamma = 2.5$. The 95\% CL ULs are $8.1\times10^{-13}$
and 
$9.6\times10^{-13}$\,cm$^{-2}$\,s$^{-1}$, respectively. For more information, refer to 
Table~\ref{tab:steadyfluxtable}. 

Given that PSR J1023+0038 shows orbital modulation in the X-ray band, a search for VHE gamma-ray emission at 
different orbital phases is also performed. The data are divided into ten phase bins and undergo the same standard 
analysis as described in Section~\ref{sec:analysis}. For both the radio MSP state and the
accretion/LMXB state, no 
significant excess is found in any of the orbital bins.

\iffalse Here lies the old table
\newcolumntype{C}{>{\centering\arraybackslash}X}
\begin{table*}
\centering
\begin{tabularx}{0.680\textwidth}{CCCC}
  \hline
  PSR J1023+0038 State & \emph{H} statistic & 95\% CL pulsed VHE flux UL (cm$^{-2}$
s$^{-1}$) & 99\% CL pulsed VHE 
flux UL (cm$^{-2}$ s$^{-1}$) \\ \hline
  Radio MSP & \boldmath$0.50$  & \boldmath$1.18 \times 10^{-12}$ & \boldmath$1.76 \times 10^{-12}$     \\ 
   Accretion/LMXB & $0.18$ & \boldmath$1.23 \times 10^{-12}$  & \boldmath$1.95 \times 10^{-12}$ \\ 
\hline
\end{tabularx}
\caption{\normalfont{{\it H} statistics and integral pulsed VHE flux upper limits computed
with the VERITAS data for both the radio MSP and accretion/LMXB states of 
PSR J1023+0038.}}
\label{tab:pulsedfluxtable}
\end{table*}
\fi

\newcolumntype{C}{>{\centering\arraybackslash}X}
\begin{table*}
\centering
\begin{tabularx}{0.680\textwidth}{CCCC}
  \hline
  PSR J1023+0038 State & \emph{H} statistic & 95\% CL pulsed VHE flux UL (cm$^{-2}$
s$^{-1}$) & 95\% CL pulsed VHE 
flux UL (flux units; erg cm$^{-2}$ s$^{-1}$) \\ \hline
  Radio MSP & $0.50$  & $1.5 \times 10^{-12}$ & $2.0 \times 10^{-12}$     \\ 
   Accretion/LMXB & - & -  & - \\ 
\hline
\end{tabularx}
\caption{\normalfont{{\it H} statistic and integral pulsed VHE flux upper limit computed
with the VERITAS data for the radio MSP state of PSR J1023+0038. Due to the unavailability of a valid pulsar timing 
solution for the accretion/LMXB state, no {\it H} statistic or flux upper limit is given for  
VERITAS data collected during this state.}}
\label{tab:pulsedfluxtable}
\end{table*}

\subsection{Search for pulsations} \label{sec:pulse}

A search for pulsed gamma-ray emission in the VHE band from the position of PSR 
J1023+0038 is performed in two parts using data recorded by VERITAS during time periods in which the radio 
MSP was active and after the disappearance of the MSP and re-emergence of an accretion disk (accretion/LMXB state).  
After applying the background-rejection and data quality cuts outlined in
Section~\ref{sec:analysis}, photon arrival times 
are barycentered and phase-folded to the pulsar period in the  
\texttt{Tempo2} software package~\citep{2006MNRAS.369..655H} using a PSR J1023+0038
Jodrell Bank ephemeris derived from radio data spanning MJD 55540 to 55644 (2010 December 10 to 
2011 March 24).  Details on the creation of the Jodrell Bank radio ephemeris can be found in 
Section 3 of~\cite{archibald2013}.  Since the pulsar ephemeris used is no longer valid during the accretion/LMXB
phase,
only results from the radio MSP state are shown.
The phase-folded light curve of PSR J1023+0038 is shown in
Figure~\ref{fig:phaseograms}. 
  De Jager's 
{\it H-test} is employed to compute {\it H} statistics that reflect the likelihood 
of the presence of a periodic signal in the light curve~\citep{1989A&A...221..180D}.  
Application of the {\it H-test} does not yield any evidence of periodicity in the VHE
gamma-ray data. 
Subsequently, integral flux limits above 
an energy threshold of 166\,GeV are computed with the method of~\cite{1994ApJ...436..239D}
assuming a 
pulsar duty cycle of 10\%, a Gaussian pulse shape, and a spectral index of 
3.8 (the same index measured by VERITAS for the Crab pulsar in~\textcite{aliu2011}). 
{\it H} statistics and integral VHE flux limits 
are given in Table~\ref{tab:pulsedfluxtable}.  
%As the phase-zero location of hypothetical gamma-ray pulsations is unknown, a test for signal
%by defining ON and OFF phase regions in the unbinned pulsar light curve is not performed.   

\begin{figure}[ht]
\epsscale{1.2}
\includegraphics[width=3.30in]{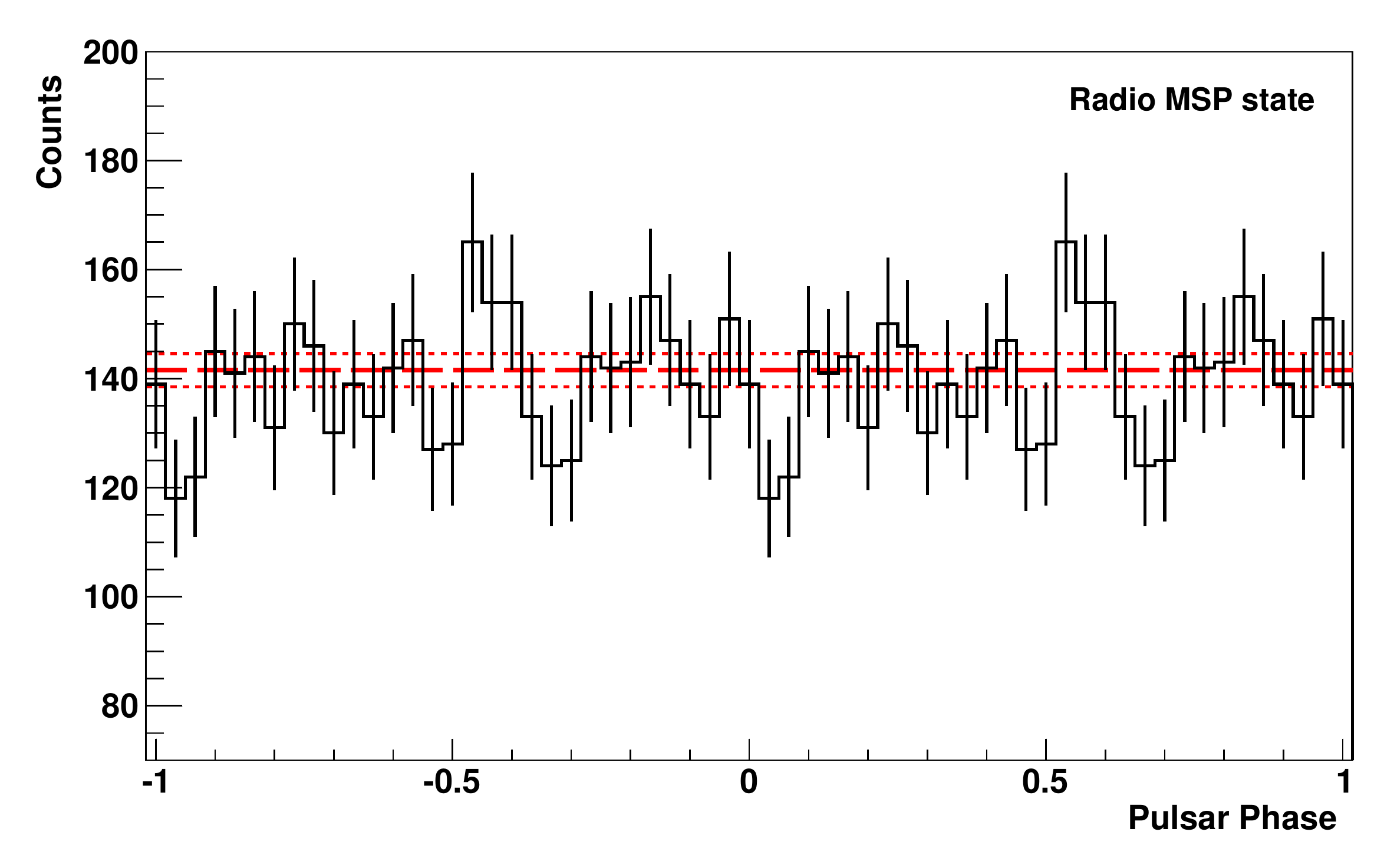}
\caption{Light curve of events phase-folded with the Jodrell Bank radio ephemeris for
the radio MSP state .  The light curve shows two pulsar periods and contains 30 bins per period.  
The dashed and dotted red lines represent the average 
number of counts and error on the average, respectively.}
\label{fig:phaseograms}
\end{figure}

\section{Discussion \& Conclusion}

During the last decade, PSR J1023+0038 has been intensively investigated in different
energy bands.
In this paper we have reported two sets of VERITAS observations, taken during the radio
MSP state and the accretion/LMXB state, that have both yielded upper limits on a VHE
gamma-ray flux.
While the beginning of the accretion phase was marked by a sharp rise of the 
luminosity both in X-rays and HE gamma rays as observed by {\it Swift} and the {\it
Fermi}-LAT~\citep{stappers2013, takata2014}, the source was not detected by VERITAS. In
the following, we discuss the constraints that can be placed on the physical properties of
PSR J1023+0038 with the VERITAS upper limits. First we will discuss the system when PSR
J1023+0038 exhibited
detectable radio pulses, and then will investigate what changed after the reappearence of
the accretion disk.

\subsection{Millisecond pulsar phase}

During the millisecond pulsar phase, radio emission from PSR J1023+0038 was characterized
by
highly frequency-dependent eclipses at superior conjunction accompanied by short,
irregular eclipses at all orbital phases~\citep{archibald2009}. Assuming a pulsar mass
$M=1.4M_\odot$ and an orbital inclination $i\sim 46^\circ$, it has been shown that the
line of sight between the pulsar and the Earth does not intersect the Roche lobe of the
companion at any point of the orbit~\citep{archibald2009}. Therefore, the eclipses must be
caused by material driven off the surface of the companion by the impinging pulsar wind.

The V magnitude of the system is orbitally modulated, reaching a minimum during
the inferior conjunction of the companion star~\citep{thorstensen2005}. Such behavior is
consistent with a Roche-lobe-filling companion near $T_{\rm eff} = 5650$\,K being
illuminated by a pulsar with an isotropic luminosity of $\sim$2$L_\odot$.

\begin{figure}[h!]
\includegraphics[width=3.30in]{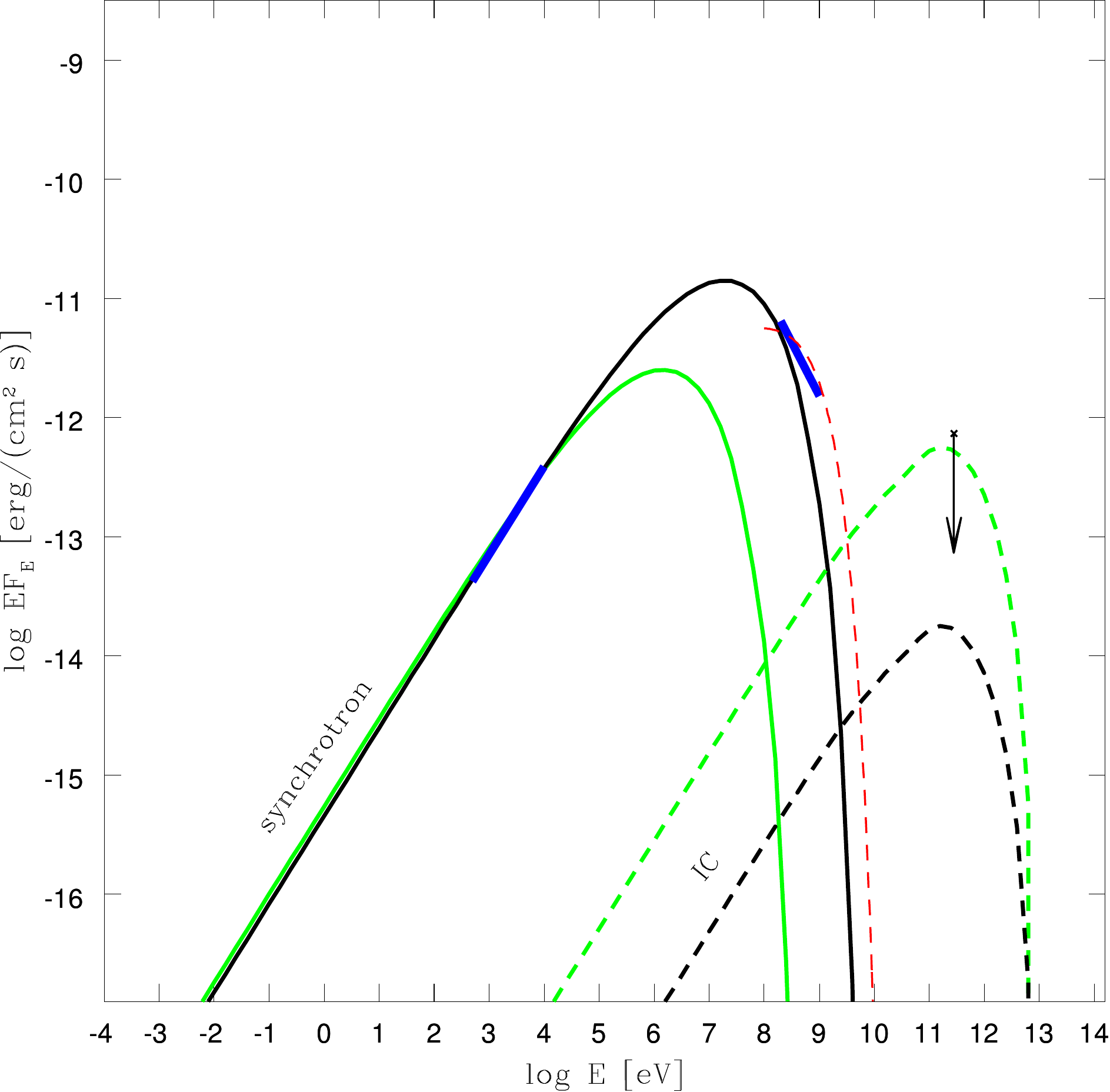}
 \caption{Broadband spectrum of PSR J1023+0038 during the millisecond pulsar phase. Thick
blue bars
show the detection of the X-ray emission by {\it XMM}-Newton in 2008~\citep{archibald2010}
and
the {\it Fermi}-LAT GeV detection~\citep{tam10}. The black solid line represents
synchrotron emission in a 40\,G magnetic field, and the black dashed line represents the
component due to inverse-Compton scattering of optical photons. The solid and dashed green lines 
represent those same components in the case of a 2\,G
magnetic field. The red dashed line represents a typical power-law model with an
exponential
cut-off spectrum of a GeV millisecond pulsar. The arrow represents the VERITAS flux upper
limit reported in this work.}
 \label{bbsp}
 \end{figure}

Orbitally modulated  X-ray emission from PSR J1023+0038 was observed by the {\it
XMM}-Newton and {\it Chandra}
X-ray observatories in 2004, 2008, and 2010~\citep{homer2006, archibald2010, bogdanov2011}. In
\cite{archibald2010}, it is shown that in the energy range 0.25 -- 2.5 keV, the X-ray
emission is also modulated at the 1.6\,ms rotational period of the MSP with a
mean-squared pulsed fraction of 0.11(2). X-ray emission observed with the {\it Swift}-XRT
in the 0.3 -- 8 keV energy range
suggests a dominant non-thermal synchrotron component originating at the intrabinary
shock. In the case of a magnetically dominated wind (with a ratio of magnetic
energy to kinetic energy $\sigma \gg 1$), the shock should
occur in a relatively strong magnetic field ($B\sim 40$\,G) due to the
small separation between the pulsar and the companion~\citep{bogdanov2011}. 
In \cite{bogdanov2011} it is shown that the depth and duration of the X-ray eclipses imply
that the intrabinary shock is localized close to the L1 Lagrangian point and has a
size of about $R\sim5\times10^{10}$\,cm.  {\it NuSTAR} has detected  a 
power-law throughout the 3 -- 79\,keV band with an estimated luminosity of  
$7.4\times 10^{32}\,\textrm{erg}\,\textrm{s}^{-1}$ 
\citep{2014ApJ...791...77T}. If the estimate of the shock size by~\cite{bogdanov2011} is correct, then a very large fraction 
of the energy in the shocked portion of the wind must be converted to X-ray emission,
which supports the high $\sigma$ scenario. 
In~\cite{archibald2010} it is also noted
that emission from the pulsar magnetosphere can contribute to the non-thermal X-ray
emission, but the orbital modulation indicates that this
component is not dominant.
 
In addition to the aforementioned non-thermal emission, there also is a faint thermal
component possibly originating from the hot polar caps of the pulsar and optically thin
thermal plasma responsible for the radio eclipses. There is no evidence in X-ray data
 for a wind
nebula associated with the pulsar. The observed X-ray luminosity in the 0.5 -- 10 keV
energy range of $L_{\rm X} \sim 10^{32}$\,erg\,s$^{-1}$~\citep[assuming a distance of
1.4\,kpc;][]{2012ApJ...756L..25D} is much
less than the spin-down
luminosity:
$L_{{\rm sd}}\simeq3.2 \times 10^{34}$\,erg\,s$^{-1}$~\citep{archibald2013}. % so what...?

The broadband spectrum of PSR J1023+0038 from X-rays to VHE gamma rays is shown in
Figure~\ref{bbsp}. While the X-ray data can be 
described by synchrotron emission from relativistic electrons exhibiting a power law with
an exponential cut-off spectral shape, $dN/dE \propto
E^{-2.52}\textrm{exp}(-E/E_{\textrm{cut}})$,  the GeV data are not readily fitted with
the same component.  However, since the X-ray
and GeV gamma-ray data are not strictly contemporaneous, spectral variability
cannot be ruled out.
The situation is similar if the observed GeV emission
is produced in the pulsar magnetosphere. The typical spectral shape of the GeV millisecond
pulsars
is a power law with an exponential cut-off, e.g.~\cite{espinoza2013}; 
see the red dashed line in Figure~\ref{bbsp} for a best fit to the {\it Fermi}-LAT
data~\citep{tam10}. This spectral shape is thought to be a result of curvature
acceleration in a gap region in the magnetosphere~\citep{harding05}. More data are
needed to distinguish between a synchrotron or curvature radiation origin of the
GeV emission, although
neither predicts emission above 10 GeV.

Synchrotron photons can inverse-Compton scatter on relativistic electrons and become
VHE photons. The ratio of the total power radiated by the synchrotron radiation and by
inverse-Compton scattering by the same distribution of electrons is equal to
$\eta=\frac{(dE/dt)_{\textrm{sync.}}}{(dE_e/dt)_{\textrm{IC}}}$.
% need to state what these variables are 
The value for
$\eta$ reaches a maximum in the Thomson limit in which
$\eta_{\textrm{T}}=\frac{B^2/8\pi}{U_{\textrm{rad}}}$, where $U_{\textrm{rad}}$ is the
energy density of the synchrotron photons. It turns out that for PSR J1023+0038, the total
energy
of
scattered photons is much smaller than the total energy of the synchrotron photons even in
the Thomson limit where 
\begin{equation}
\eta_\textrm{T} \sim 600 \frac{(B/40{\rm G})^2(R/5\times
10^{10} {\rm cm})^2 }{L/10^{32}{\rm erg/s}}. 
\label{etaT}
\end{equation}

An additional potential source of VHE emission is external inverse-Compton scattering of soft
photons from the optically bright companion with an effective temperature of
$T=5650\,K$~\citep{thorstensen2005}. This inverse-Compton component is shown in 
Figure~\ref{bbsp} as a black dashed
line. Given the assumed value of the magnetic field, $B=40\,{\rm G}$, the component lies
well below the VHE flux upper limit. However, for a lower magnetic
field strength, the difference between the peak flux values of the synchrotron and
inverse-Compton components will become smaller, allowing VERITAS observations to set a
lower limit on the magnetic field strength.  As shown by
the green lines in Figure~\ref{bbsp}, the case of a 2\,G magnetic field gives close to the
peak inverse-Compton flux allowed by the upper limit derived from the VERITAS data.

Note that for a 2\,G  magnetic field, $\eta_\textrm{T}$ defined by 
Equation~\ref{etaT} is close to unity. However, X-ray photons will be up-scattered in the Klein-Nishina 
regime, and in this case the total energy of scattered photons is much smaller than in 
the Thomson regime:
\begin{equation}
\eta_{\textrm{KN}}=\frac{B^2/8\pi}{\frac{9}{32}U_{\textrm{rad}}}\frac{\textrm{ln} (\frac {\hbar
\omega_0 \gamma}{m_ec^2}) }{\gamma^2 \hbar^2 \omega_0^2 / (m_ec^2)^2}\sim
\eta_\textrm{T}/2000 
\end{equation}
for 1 keV photons scattered into the VHE band by electrons with
$\gamma=10^4$~\citep{2011hea..book.....L}. Although lower-energy photons are scattered in the transition regime between the  
Thomson and Klein-Nishina regimes, their energy density is much lower than that of the X-ray
photons, and so the self-scattering process is not important in this case either.

%\textbf {Scattering of photons with lower energy happens  in the transition regime between Thomson and Klein-Nishina, but their  energy density is much lower  than that of the X-ray
%photons, and so the self scattering process is not important in this case either.}

 %%%%%%%%%%%%%%%%%%%%
 %%%%%%%%%%%%%%%%%%%%
\begin{figure}[h]
\includegraphics[width=3.30in]{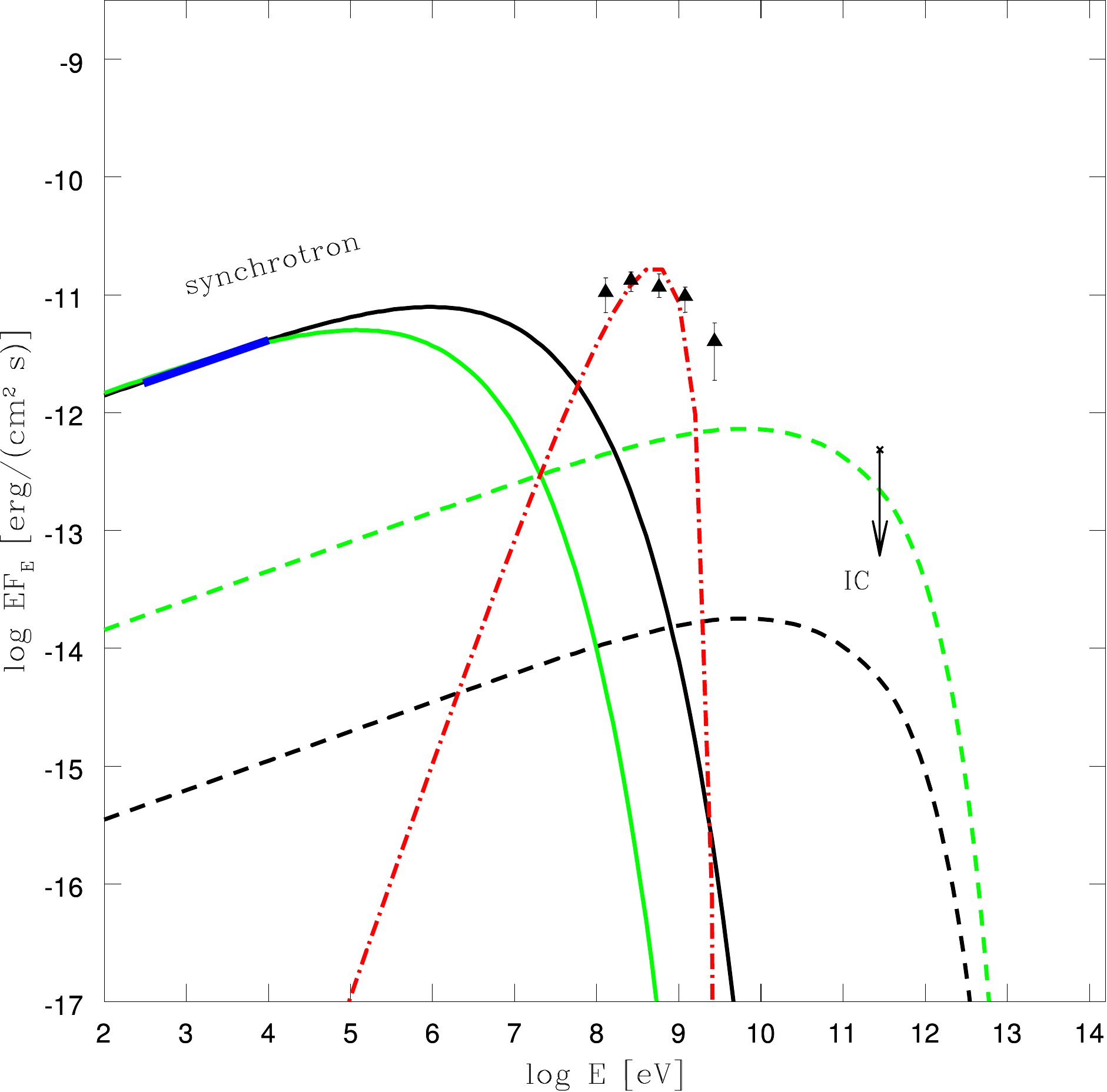}
 \caption{Broadband spectrum of PSR J1023+0038 after the reappearance of the accretion
disk.
The thick
blue bar shows the X-ray emission detected by {\it Swift} in 2013 November
\citep{takata2014}. Black
triangles represent the {\it Fermi}-LAT HE gamma-ray detection in
2013~\citep{takata2014}.  The arrow shows the
VERITAS upper limit for the accretion/LMXB state, as reported in this work. Solid and
dashed lines correspond to the synchrotron and inverse-Compton emission coming from the
shock for the
case of a 10\,G (green lines) and 80\,G (black lines) magnetic field. The spectral
signature of inverse-Compton scattering of photons emitted by the accretion disk
on the unshocked electrons is shown with a red dash-dotted line.}
 \label{acr}
 \end{figure}
 %%%%%%%%%%%%%%%%%%%%
 %%%%%%%%%%%%%%%%%%%%

\subsection{Accretion phase}

The reappearance of the accretion disk in 2013 June was accompanied by the disappearance
of
radio pulsations and an increase of the X-ray and HE gamma-ray luminosities.
Accreting binary systems are not typically bright in the GeV
domain. The only two binaries detected by the {\it Fermi}-LAT in which the presence of an
accretion disk is certain are Cyg X-3 and Cyg X-1~\citep{corbel2012,malyshev2013}, and in
both cases the HE emission is not believed to come from the disk, but rather to be
generated in the relativistic jet. The formation of a jet in PSR
J1023+0038 has not
been
observed in VLBI imaging, although variable point-source emission has been
seen~\citep{deller2015}. 
Further, it appears that the X-ray pulsations, indicating accretion onto the neutron star
surface, are intermittent~\citep{archibald2014}. Therefore it could be the case that, as
discussed by~\cite{takata2014,coti2014,li2014,papitto2015}, the rotation-powered 
MSP is still at least partially active in PSR J1023+0038, and the complete disappearance
of the
pulsations is due to absorption by matter evaporating from the accretion disk. In this
case, the principal differences from the radio MSP state discussed in the previous section
would be a) the presence of additional soft photons emitted by the accretion disk and b)
the shift of the shock closer to the pulsar due to the inward pressure of the disk.
% increased mass-loss rate of the companion. 

The presence of additional photons from the accretion disk leads to an increase of the HE
luminosity as a result of scattering of those photons on the unshocked electrons of the
pulsar wind~\citep{takata2014}. The result of the scattering of the UV photons with
temperature $T=10$\,eV on the cold relativistic electrons with Lorentz factor
$\gamma=10^4$ is shown in Figure~\ref{acr} with a red dash-dotted line. The shift of the
shock closer to the pulsar up to a distance $r=5 \times 10^{10}$\,cm~\citep{takata2014}
will lead to the increase of the magnetic field by a factor of two in comparison to the
field strength discussed in the previous section if the magnetic field is dominated by
that in the pulsar wind. The resulting synchrotron and inverse-Compton emission from the
shocked electrons generated in the region with $B=80$\,G is shown in Figure~\ref{acr} with
black solid and dashed lines, respectively. The VERITAS upper limit clearly shows that
the field in the region cannot be much smaller than 10\,G (green lines in
Figure~\ref{acr}).  Thus the VERITAS observations before and after the source state change
put limits on the minimum value of the magnetic field in a compact, synchrotron-emitting
region, regardless of the precise mechanism of the charge acceleration or the source of
the magnetic field.

We note that~\cite{papitto2015} have proposed pulsar magnetic 
field threading of the accretion disk down to the corotation radius of PSR J1023+0038 ($\sim$\,24\,km) 
as a field source for the synchrotron emission. Were this the case, the strength of the magnetic 
field could be much larger.

The VERITAS limits support the conclusion of the magnetically-dominated pulsar wind 
discussed in~\cite{bogdanov2011}. However, in both the MSP and accretion/LMXB states,
there are alternative sources of magnetic fields that should be considered, namely that of
the companion in both cases and that of the accretion disk itself in the second case. 
Assuming that the companion is tidally locked, observations of rapidly-rotating, low-mass
stars suggest a surface magnetic field strength on the order of 1\,kG~\citep{morin2012}.
The observed orbital variations may also indicate a strong,
subsurface magnetic field in the companion~\citep{archibald2013}.

\acknowledgements
The authors are grateful to A. Archibald for providing a Jodrell Bank radio ephemeris for
PSR J1023+0038, which made the analysis described in Section~\ref{sec:pulse} possible.

This research is supported by grants from the U.S. Department of Energy Office of Science,
the U.S. National Science Foundation and the Smithsonian Institution, and by NSERC in
Canada. We acknowledge the excellent work of the technical support staff at the Fred
Lawrence Whipple Observatory and at the collaborating institutions in the construction and
operation of the instrument. Ester Aliu acknowledges support by the Spanish Ministerio de
Economia y Competitividad (MINECO) under grants AYA2013-47447-C3-1-P. The VERITAS
Collaboration is grateful to Trevor Weekes for his seminal contributions and leadership in
the field of VHE gamma-ray astrophysics, which made this study possible.

{\it Facility:} \facility{VERITAS}.

\end{document}